# Análise dinâmica da tendência para o equilíbrio num modelo simples: a 2ª Lei de Newton $f = ma$ e a 2ª Lei da Termodinâmica $dS \geq 0$

Rodrigo de Abreu
Departamento de Física e Centro de Electrodinâmica,
Instituto Superior Técnico, Lisboa, Portugal

**Abstract**

We relate Newton's Second Law with the Second Law of Thermodynamics through the analysis of a simple model introducing a dynamic pressure concept. From this analyses we can clarify some conceptual problems resulting from several concepts of heat and work on the First Law of Thermodynamics.

**1. Introdução**

A tendência para o equilíbrio num modelo unidimensional foi anteriormente estabelecida [1]. O modelo então utilizado e ao qual novamente nos referimos, é constituído por um gás ideal clássico. As partículas do gás deslocam-se perpendicularmente a um êmbolo, que confina o gás. O gás e o êmbolo estão inseridos num cilindro cujas paredes reflectem as partículas elasticamente. O êmbolo move-se devido á aceleração da gravidade e á acção devida ás colisões das partículas. As partículas movem-se sem exercerem qualquer força de interacção entre si movendo-se paralelamente umas às outras, apenas colidindo com o êmbolo e com o fundo do cilindro. É nestas colisões com o êmbolo que emerge o mecanismo de interacção origem do atrito interno, que leva o êmbolo á posição final de equilíbrio [1]. Movendo-se as partículas com uma velocidade muito superior à velocidade do êmbolo verifica-se, por unidade de tempo, um elevado numero de colisões das partículas com o êmbolo. Estas colisões originam uma pressão. Quando o êmbolo está em movimento esta pressão é uma pressão dinâmica. De facto, se o êmbolo se deslocar no sentido do aumento de volume com uma velocidade diferente de zero, a pressão é ligeiramente inferior à pressão estática, a pressão sobre o êmbolo em repouso. Se o êmbolo comprimir o gás a pressão dinâmica é ligeiramente superior à pressão estática. O movimento do êmbolo submetido a esta pressão dinâmica e ao peso determina-se a partir da 2ª Lei de Newton. Esta equação do movimento tem um ponto assimptóticamente estável, o ponto de equilíbrio do êmbolo [1]. Deste resultado, a tendência para o equilíbrio obtida por aplicação da 2ª Lei de Newton, conclui-se que a energia interna para um mesmo volume (quando o êmbolo regressa a um determinado volume) é superior ao valor anteriormente assumido para esse mesmo volume [2]: a energia mecânica do êmbolo degrada-se em energia interna, $U$, do gás. Deste facto podemos escrever $U=U(V,S)$, $dU \geq 0$, $dS \geq 0$. Somos conduzidos dessa forma à Segunda Lei da Termodinâmica, isto é, a variação de entropia, $S$, de um Sistema (conjunto de sub-sistemas e portanto térmicamente isolado, no caso presente o gás) é positiva ou nula [2]. A variação de entropia nula define a transformação reversível, que corresponde ao limite de um movimento extremamente lento, em equilíbrio, em que globalmente se pode regressar a valores anteriormente assumidos - quando o êmbolo regressa a uma posição anterior a energia interna regressa ao valor anterior, não havendo variação de entropia.

No presente artigo, e tendo como base o mesmo modelo simplificado inicialmente referido, pretende-se estabelecer a ligação entre a descrição dinâmica permitida pela 2ª lei de Newton e as equações da termodinâmica, em particular a equação da variação da entropia. Na secção 2 mostra-se como a distinção entre pressão estática e pressão dinâmica conduz à necessidade de introduzir uma grandeza que se identifica com a entropia tradicional da Termodinâmica, concluindo-se que durante o regime dinâmico essa quantidade varia monotònicamente definindo a tendência para o equilíbrio. Na secção 3 obtém-se a equação dos gases ideais clássicos ("gases perfeitos") e determina-se a entropia em função da temperatura e do volume. Na secção 4 estabelece-se a equação da variação da entropia ao longo do tempo, a qual será utilizada na simulação do regime dinâmico apresentada no Apêndice I. Mostra-se ainda como os resultados obtidos no modelo simplificado permitem esclarecer as condições de validade de certas aproximações por vezes utilizadas na análise termodinâmica de situações mais complexas. No Apêndice I deduz-se a equação do regime dinâmico do sistema por aplicação da 2ª lei de Newton, considerando que o movimento do êmbolo é determinado pela pressão dinâmica e pela acção gravítica, e apresentam-se os resultados da simulação numérica que evidenciam a tendência para o equilíbrio. No Apêndice II analisam-se as condições de validade da aproximação da pressão dinâmica pela pressão estática e as consequências desta aproximação na formulação do 1º Princípio da Termodinâmica.

**2. A 2ª Lei de Newton, a pressão dinâmica e a tendência para o equilíbrio.**

Consideremos a fig. 1:

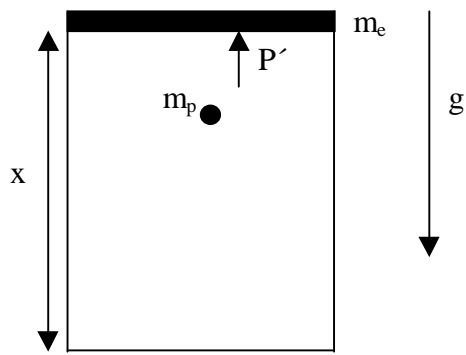

**Fig. 1**
A pressão $p'$ é a pressão dinâmica sobre
o êmbolo diferente da pressão estática $p$
correspondente à pressão que o gás exerceria
sobre o êmbolo em repouso. A área do êmbolo
é unitária e portanto o volume V do gás é nume-
ricamente igual a $x$, a altura do embolo. A massa do
êmbolo é $m_e$ e a aceleração da gravidade é $g$. No-
te-se que se existisse gás do outro lado do êmbolo
a pressão sobre o êmbolo seria a diferença das pres-
sões (estáticas ou dinâmicas).

A 2ª Lei de Newton permite escrever

$$m_e \ddot{x} = p' - m_e g \qquad (1)$$

Dado que a pressão dinâmica $p'$ sobre o êmbolo depende do sentido do movimento do êmbolo, sendo superior ou inferior à pressão estática $p$ conforme o êmbolo se desloque "contra" o gás ou "a favor" do gás, o êmbolo vai mover-se trocando energia com as partículas do gás de uma forma assimétrica (este processo é objecto de um tratamento quantitativo com simulação numérica no Apêndice I). Para compreender qualitativamente que assim é, basta notar o seguinte:

1. A energia das partículas do gás é essencialmente cinética (sem perda de generalidade desprezamos a energia potencial gravitacional) [1]. A soma das energias cinéticas das partículas é a energia interna do gás, $U$.

2. A energia que o êmbolo troca com o gás ao longo do movimento, vai-se distribuindo pelas diversas partículas, podendo distribuir-se de múltiplas maneiras dependentes da distribuição no tempo e no espaço que define colectivamente o movimento do gás [1].

3. Quando o êmbolo está num movimento ascendente (contrário à aceleração g) a pressão dinâmica é inferior à pressão estática que se exerceria sobre o êmbolo se este estivesse em repouso. Quando o êmbolo está num movimento descendente (no sentido da aceleração g) a pressão dinâmica é superior à pressão estática que se exerceria sobre o êmbolo se este estivesse em repouso. Portanto a energia que o gás cede ao êmbolo é inferior à energia que o êmbolo cede ao gás. Desta forma a amplitude do movimento do êmbolo vai diminuindo até atingir o repouso (dado o anteriormente afirmado, o êmbolo oscilaria indefinidamente se a pressão sobre o êmbolo em movimento fosse a pressão estática).

Podemos analiticamente exprimir 1., 2., 3., através do Teorema das Forças Vivas (TFV). O trabalho das forças $dW$ que actuam sobre o êmbolo é a soma do trabalho do peso e da pressão dinâmica do gás, respectivamente $dW_g$ e $dW_{p'}$. Este trabalho é (TFV) igual à variação da energia cinética do êmbolo

$$dW = dEcin = dW_g + dW_{p'}.$$

Do princípio de conservação de energia temos que $U + Ecin + Epot = Constante$, em que $U$ é a energia do gás, $Ecin$ a energia cinética do êmbolo e $Epot$ a energia potencial gravitacional do êmbolo. O trabalho da força peso, $dW_g$, é simétrico da variação da energia potencial $dE_{pot}$. Temos portanto

$$dEcin + dEpot + dU = 0, \qquad (2)$$

$$dW = dEcin, \qquad (3)$$

$$dWg = -dEpot, \qquad (4)$$

$$dW - dW_g + dU = 0, \qquad (5)$$

$$dW_g + dW_p' - dW_g + dU = 0, \qquad (6)$$

$$dW_p' = -dU, \qquad (7)$$

isto é, o trabalho da pressão dinâmica é igual ao simétrico da variação da energia do gás. Podemos exprimir este resultado na forma

$$dU = -p'dV \qquad (9)$$

em que $V$ é o volume ocupado pelo gás.

Dado que a pressão dinâmica $p'$ é superior ou inferior à pressão estática $p$, dependendo da variação de volume ser negativa ou positiva, temos que

$$dU = -p'dV \geq -pdV. \qquad (10)$$

A variação da energia do gás só é igual a $-pdV$ no limite em que a velocidade do êmbolo é zero. Tal condição verifica-se quando, a partir duma situação de equilíbrio, em que o peso do êmbolo é igual à pressão estática do gás e o êmbolo não tem energia cinética, se altera progressivamente o peso do êmbolo através do acréscimo de massas infinitesimais, "grãos de areia". Da relação anterior (10) temos que

$$dU = -pdV + \text{\textit{quantidade infinitesimal positiva}} \atop \text{\textit{ou nula}}. \qquad (11)$$

Podemos analiticamente exprimir este resultado introduzindo uma variável $S$ (na energia interna) [2]

$$U = U(V,S). \qquad (12)$$

De (12) temos

$$dU = (\partial U/\partial V)_S \, dV + (\partial U/\partial S)_V \, dS. \qquad (13)$$

Façamos

$$(\partial U/\partial S)_V = \lambda, \qquad (14)$$

e arbitremos $\lambda > 0$. Esta arbitrariedade acarreta que durante o regime dinâmico $dS \geq 0$, dado (11) e dado que

$$(\partial U/\partial V)_S = -p. \qquad (15)$$

A variável $S$ assim introduzida identifica-se com a tradicional grandeza entropia da Termodinâmica.

À medida que o êmbolo se movimenta o gás vai aumentando de entropia e a amplitude do movimento do êmbolo vai diminuindo até se atingir o repouso. A entropia atinge o máximo compatível com o peso associado ao êmbolo.

### 3. Determinação do estado final de equilíbrio: cálculo da pressão, do volume, de $\lambda$ e de $S$.

A pressão que o gás exerce sobre o êmbolo em repouso pode facilmente ser calculada. A variação da quantidade de movimento de uma partícula numa colisão com o êmbolo é igual a $2\,m_p\,v$ (consideremos por simplicidade e sem perda de generalidade que a partícula se movimenta perpendicularmente ao êmbolo com velocidade $v$). Dado que a frequência com que uma partícula colide com o êmbolo é o inverso do tempo entre colisões, $2\,x/v$, isto é $v/(2\,x)$, temos que a variação da quantidade de movimento por unidade de tempo é $2\,m_p\,v\,v/(2\,x)$, $(2/2)\,m_p\,v^2/x$. Para $N$ partículas a pressão é evidentemente

$$p = N\,2\,(1/2)\,m_p\,v^2/x, \qquad (16)$$

ou seja a pressão de um gás de partículas é

$$p = 2\,U/V, \qquad (17)$$

dado $U = N\,(1/2)\,m_p\,v^2$ e $x = V$, dado considerarmos a área do êmbolo unitária. Não tendo as partículas todas a mesma velocidade há que calcular o valor médio de $v^2$, isto é passamos a ter $U = N\,(1/2)\,m_p\,<v^2>$, e se as partículas se moverem em três dimensões a pressão passa a ser evidentemente $p = 2/3\,U/V$.

Admitamos que no início o êmbolo de massa $m_e$, se encontra com velocidade nula e que a energia do gás é $U_0$. Após o êmbolo dissipar a diferença de energia potencial entre a altura inicial e final, temos

$$-m_e\,g\,\Delta x = \Delta U = \tfrac{1}{2}\,\Delta(p\,V), \qquad (18)$$

ou

$$-m_e\,g\,(V - V_0) = \tfrac{1}{2}\,(pV - p_0 V_0). \qquad (19)$$

Dado que no equilíbrio

$$p = m_e\,g, \qquad (20)$$

as equações (19) e (20) permitem calcular a pressão final de equilíbrio $p$ e o volume final de equilíbrio $V$ (*ver apêndice II*)

Para determinar os valores de $\lambda$ e de $S$, partamos das relações (14) e (15). De $p = 2\,U/V$ temos derivando ambos os membros em ordem a S, e dado (14),

$$(\partial p/\partial S)_V = 2\,(\partial U/\partial S)_V/V = 2\,\lambda/V. \qquad (21)$$

De (15) temos derivando ambos os membros em ordem a $S$

$$(\partial^2 U/\partial S \partial V) = -(\partial p/\partial S)_V \qquad (22)$$

De (14) temos derivando ambos os membros em ordem a *V*

$$(\partial^2 U/\partial S \partial V) = (\partial \lambda/\partial V)_S \qquad (23)$$

ou seja, comparando (22) e (23)

$$(\partial \lambda/\partial V)_S = -(\partial p/\partial S)_V, \qquad (24)$$

e dado (21)

$$(\partial \lambda/\partial V)_S = -2 \lambda/V. \qquad (25)$$

De (25) concluímos que ao longo de uma transformação em que *S* não varia,

$$d\lambda/\lambda = -2\, dV/V. \qquad (26)$$

Integrando (26) temos que ao longo de uma isentrópica ($dS = 0$)

$$\lambda V^2 = Const.. \qquad (27)$$

De $p = -(\partial U/\partial V)_S = 2\, U/V$, (15) e (17), obtemos da mesma forma

$$UV^2 = Const.. \qquad (28)$$

Comparando (27) e (28) obtemos, qualquer que seja S

$$U = A\, \lambda \qquad (29)$$

em que *A* é constante para um dado valor de *S*. Podemos portanto escrever, de (17) e (29),

$$p = 2\, U/V = 2\, A\, \lambda/V \qquad (30)$$

ou seja

$$pV = B\, \lambda \qquad (31)$$

em que $B = 2A$ é constante ao longo de uma transformação isentrópica ($dS=0$).

Se se admitir que *B* é constante, o que constitui uma boa aproximação para um gás ideal clássico [3], podemos calibrar $\lambda$. Foi o que empiricamente se admitiu quando se construiu o primeiro termómetro de ar a pressão constante, em que se admitiu uma linearidade entre *V* e *a temperatura empírica* (para o ar nas condições *C.N.T.P. B* é aproximadamente constante quando se muda de isentrópica, dado o ar estar naquelas condições a comportar-se de acordo com a estatística de Maxwell-Boltzmann como um gás ideal clássico, para a qual a energia do gás é proporcional ao parâmetro $\lambda$). Desta forma temos que o parâmetro $\lambda$ se pode identificar com a temperatura medida por um termómetro, dado o acordo entre a não variação de $B$ e a hipótese de uma relação linear entre a temperatura e o volume.

Com esta hipótese, a de que a energia é proporcional a $\lambda$, pode-se medir a constante $B$ e escrever $B = NK$ em que $N$ é o numero de partículas e $K$ é uma constante. Se fizermos $\lambda=T$ (em que T é a temperatura) temos que nesta nova notação (31) passa a ser

$$pV = N\,K\,T. \tag{32}$$

Tendo os valores de $N$, de $K$, sabendo $p$ e $V$ ( de (19) e (20)) sabemos o valor de $\lambda$, o valor de $T$.

A determinação da variação de $S$ faz-se com o recurso à equação (13), que agora se pode escrever, atendendo a (15), (14) e que $\lambda =T$:

$$dU = -p\,dV + T\,dS. \tag{33}$$

De (33) temos

$$dS = (dU + pdV)/T \tag{34}$$

e como de (30) e (32):

$$p = 2\,U\,/V = 2\,A\,T/V = N\,K\,T/V, \tag{35}$$

temos

$$A = NK/2, \tag{36}$$

$$U = N\,K\,T\,/\,2, \tag{37}$$

$$dS = (1/2)\,N\,K\,dT\,/\,T + N\,K\,dV\,/\,V. \tag{38}$$

Integrando (38) temos

$$S = S_0 + (1\,/\,2)\,N\,K\,Ln\,(\,T\,/\,T_0) + N\,K\,Ln\,(\,V\,/\,V_0). \tag{39}$$

### *4. A variação de entropia em ordem ao tempo*

De (9) e (33) temos

$$dU = -\,p\,´dV = -\,p\,dV + T\,dS \tag{40}$$

Durante o movimento do êmbolo temos, de (40)

$$T\,dS = -(p' - p)\,dV \tag{41}$$

ou

$$\dot{S} = -(p' - p)\,\dot{x}/T \tag{42}$$

em que $\dot{S}$ e $\dot{x}$ são a derivada primeira em ordem ao tempo de $S$ e de $x$. Para se calcular $\dot{S}$ é necessário conhecer-se $p'$ em cada instante. De facto se $p'$ for conhecido pode determinar-se, $dx$ através da equação (1), $dU$ através da equação (9), $p$ através da equação (17), $T$ através da equação (32) e finalmente $\dot{S}$ (*ver apêndice I*).

    A pressão dinâmica $p'$ só em modelos simples como o de um gás ideal clássico [1, 3] ou um gás de fotões [4] é que pode ser facilmente determinada em função de $p$. Mas conceptualmente, a descrição que acabamos de fazer é geral e verifica-se para todos os regimes dinâmicos. Exemplifiquemos com o movimento de uma massa $m$ a oscilar submetida a acção do peso e à acção da força de uma mola. Em rigor, a força exercida pela mola depende da velocidade da massa que nela se pendurou. A solução correspondente ao movimento oscilatório harmónico linear sem atrito corresponde a se estar a admitir que a força dinâmica é igual à força estática ($dS = 0$), e a admitir que os parâmetros da mola nessa transformação isentrópica, por exemplo $T$, não variam.

    Embora não seja possível, com generalidade, determinar $p'$, a abordagem do problema simplificado baseado no modelo unidimensional, dá indicação clara do que é possível determinar em situações complexas. Através da determinação experimental de coeficientes de variação $\beta = (1/V)\,(\partial V/\partial T)_p$ e $k = -(1/V)\,(\partial V/\partial p)_T$ é possível, integrando, determinar a equação de estado $p = p(T,V)$. Através da determinação experimental de $C_V = (\partial U/\partial T)_V$, para um determinado volume $V_0$, $C_{V0}$, determina-se $dS = (\partial S/\partial T)_V\,dT + (\partial S/\partial V)_T\,dS$, dado $(\partial S/\partial T)_V = (C_V/T)$ e $(\partial S/\partial V)_T = (\partial p/\partial T)_V$, como é bem conhecido. Desta forma é possível determinar $U = U(T,V)$. Desta forma é possível, por exemplo, determinar a situação final de equilíbrio se se souber a descrição no tempo da força exterior (por exemplo o peso constante do êmbolo) (*ver apêndice II*). O trabalho da força exterior entre dois pontos de equilíbrio é igual ao trabalho da força dinâmica entre os mesmos pontos. Esta é uma situação complexa que se pode resolver.

    Uma outra aproximação que em determinadas situações pode ser feita é aproximar $p'$ por $p$, o que equivale a desprezar o termo $TdS$ na equação (33) em face de $-pdV$ (*ver apêndice II*), *transformação isentrópica*, ou considerando que $p'$ é $p$, considerar que $TdS$ é a energia trocada com o exterior diferente da parcela $-pdV$. Mas o que não é aceitável, nem como uma aproximação, é considerar que o trabalho da força exterior é igual ao trabalho da força interior mesmo que o êmbolo esteja em movimento muito lento, bastando para tal notar, que quando o êmbolo é solto, no exemplo que temos vindo a considerar, a pressão exterior devida ao peso do êmbolo é diferente da pressão interior, embora no arranque as pressões dinâmicas e estática sejam praticamente iguais e o trabalho da força interior possa ser aproximado por $-pdV$ (*ver apêndice II*). Pode acontecer que, durante parte da trajectória dinâmica se verifique que o trabalho da força interior é aproximadamente dado pelo integral de $-pdV$, dado que o integral de $TdS$ ainda é desprezável (*ver apêndice II*) sem no entanto se verificar igualdade entre os trabalhos elementares da força interior e da força peso. Em transformações reversíveis, contudo, o trabalho da força exterior é, sempre, igual ao trabalho da força interior, dado o êmbolo estar virtualmente sempre em repouso, isto é dado a força exterior ser sempre a menos de uma quantidade infinitesimal igual à força interior ($dS = 0$). Numa

transformação em que se realiza trabalho através de um dispositivo como o agitador de pás mecânicas de *Joule* (*paddle-wheel experiment*) [5], o trabalho evidentemente deixa de ser dado por –*p´dV* (embora eventualmente uma parcela do trabalho possa ser dada por –*p´dV* se existir um êmbolo que se desloque submetido a uma força peso).

Num regime dinâmico, em que há aceleração, só como aproximação é que a força dinâmica é a força estática. A transformação rigorosamente reversível é uma transformação rigorosamente isentrópica. Só como aproximação é que um regime dinâmico é isentrópico e a afirmação da reversibilidade das equações da mecânica contém esta aproximação: se se despreza o afastamento da força dinâmica da força estática obtêm-se diversos valores da energia do gás para os diversos valores da variável de deformação, no caso analisado o volume, sem que a entropia varie. Estes valores são os que rigorosamente seriam atingidos, através da alteração infinitesimal da força exterior ("grãos de areia") em relação à força interior, isto é sem qualquer aceleração. Por isso é que a trajectória dinâmica, com esta aproximação, é dita reversível [6, 7], dado que faz uma descrição no tempo sem que a entropia varie – o regime é periódico, o gás e o êmbolo regressam periodicamente às condições iniciais, a energia do gás comporta-se como se fosse uma função potencial. Pode neste caso e nos similares fazer-se uma abordagem variacional - a abordagem vectorial de Newton e a abordagem da teoria variacional de Euler e Lagrange são equivalentes, existe uma "work function", a entropia não varia [8, 9]. Temos assim duas ordens de aproximação – a "mecânica", cuja descrição isentrópica, resulta do valor médio (em si uma aproximação) aproximadamente calculado por não considerar o movimento do êmbolo, e que se baseia na pressão estática - a "termodinâmica", que contendo a aproximação inerente ao valor médio corrige-o ao considerar o movimento do êmbolo, dando significado à pressão dinâmica.

As idealizações têm de ser rigorosamente construídas. As aproximações têm de ser suportadas nas idealizações, e têm de ser aplicadas com cautela, pois podem dar origem a interpretações contraditórias ou a restrições desnecessárias [6, 7, 10, 11, 12-26]: a grandeza com a forma *dQ* que aparece na equação do 1º Princípio pode não corresponder a "calor trocado com o exterior" (ver apêndice II). No entanto é possível generalizar as equações para transformações irreversíveis que não sejam "quase-estáticas" (transformações em que a pressão exterior é igual à pressão estática [10], ou numa outra acepção transformações feitas ao longo de pontos de equilíbrio Termodinâmico [6, 7, 10,11, 12-26]), bastando para tal que se introduza o conceito de pressão dinâmica.

**Conclusão**

Através da análise de um modelo simplificado estabeleceu-se a ligação entre a descrição dinâmica permitida pela *2ª Lei de Newton* e as equações que resultam da introdução da variável entropia - a *2ª Lei da Termodinâmica*. A variável temperatura surge como grandeza derivada permitindo resolver situações complexas que, com generalidade, só poderiam ser resolvidas pela *2ª Lei de Newton* se fosse possível determinar a força dinâmica que só como aproximação é a força estática. O modelo apresentado, dada a simplicidade, permite de uma forma clara compreender a origem, significado e condições de validade de algumas aproximações que se fazem na Termodinâmica ou eliminar restrições desnecessárias.

**Apêndice I**

Consideremos uma simulação do gás através da substituição das $N$ partículas do gás por uma só partícula (*ver apêndice II*) cuja massa é igual à massa das $N$ partículas e com a mesma energia do gás [1]. É razoável supor que esta é uma boa aproximação para massas do êmbolo muito superiores à massa do gás e em que a velocidade da partícula é muito elevada comparada com a velocidade do êmbolo, e para valores de $x$ pequenos. Nestas condições o numero de colisões na unidade de tempo é muito elevado.

Se nos colocarmos no referencial do êmbolo vemos o fundo do cilindro deslocar-se com uma componente de velocidade $\dot{x}$ e a componente da velocidade da partícula passa a ser $(u - \dot{x})$. A pressão $p'$ obtém-se substituindo em (16) $v$ por $(u - \dot{x})$ e $N = 1$

$$p' = m_p (u - \dot{x})^2 / x \tag{1}$$

ou

$$p' = m_p u^2 (1 - \dot{x}/u)^2 / x \tag{2}$$

que para $|\dot{x}| << |u|$ tem o valor aproximado

$$p' = m_p u^2 (1 - 2\dot{x}/u)/x \tag{3}$$

A equação de conservação da energia escreve-se

$$E_0 = (m_p + m_e) g x_0 + m_p u_0^2/2 + m_e \dot{x}_0^2/2 = \\ = (m_p + m_e) g x + m_p u^2/2 + m_e \dot{x}^2/2 \tag{4}$$

A 2ª Lei de Newton escreve-se

$$m_e \ddot{x} = p' - m_e g$$

ou

$$\ddot{x} = p'/m_e - g \tag{5}$$

Inserindo (3) em (5) vem

$$\ddot{x} = (m_p/m_e) u^2 (1 - 2\dot{x}/u)/x - g \tag{6}$$

De (4) temos

$$2 E_0/m_e - (1 + m_p/m_e) g x - \dot{x} = (m_p/m_e) u^2 \tag{7}$$

Substituindo (7) em (6) temos

$$\ddot{x} = \{2 E_0/ m_e - (1 + m_p/ m_e) g x - \dot{x}^2\}(1 - 2 \dot{x}/u)/x - g \qquad (8)$$

ou

$$\ddot{x} = \{2 E_0/ m_e - (1 + a) g x - \dot{x}^2\}(1 - 2 \dot{x}/u)/x - g \qquad (9)$$

em que $a = m_p/ m_e$.

Esta equação pode facilmente ser resolvida com o recurso ao programa *Mathematica*.

Seguidamente apresenta-se o programa que permite representar graficamente $(x, \dot{x})$, $(x,t)$ e $(\dot{x},t)$. Também se representa gráficamente $(\dot{s},t)$, calculada a partir da equação (42) obtida anteriormente na secção 4.

*en=(1+a)g x0+ a u0^2/2 + x´0^2/2;*
*g=1;*
*a=10^(-4);*
*x´0=0;*
*u0=100;*
*x0=2;*
*en*
 *u[t]=Sqrt[(2 en - 2 (1+a) g x[t] - y[t]^2)/a];*
*soln=NDSolve[*
*{x'[t]==y[t],*
*y'[t]==((1/x[t])(2 en - 2 (1+a) g x[t] -*
*(y[t]^2))(1-(2 y[t])/u[t])-g),x[0]==2.0,*
*y[0]==0},{x,y},{t,50},MaxSteps->2000]*
*ParametricPlot[Evaluate[{x[t],y[t]}/.soln],{t,0,50}, PlotRange->All,*
*PlotPoints->50]*
*Plot[Evaluate[{x[t],y[t]}/.soln],{t,0,50}, PlotRange->All,*
*PlotPoints->50]*

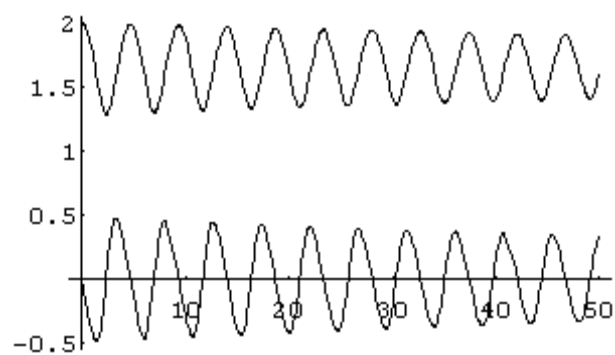

Fig. 2 *Representação da posição e da componente da velocidade do êmbolo em função do tempo.*

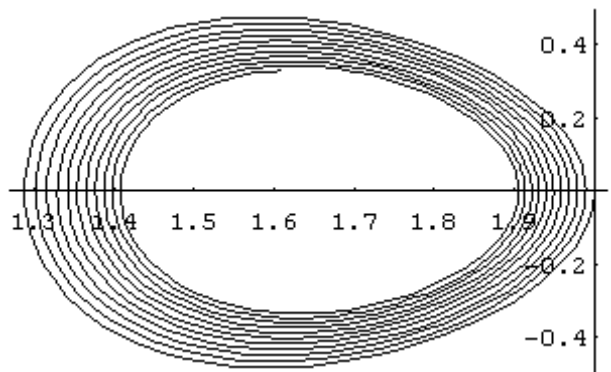

Fig. 3 *Representação no espaço de fase do movimento do êmbolo (x, $\dot{x}$ )*.

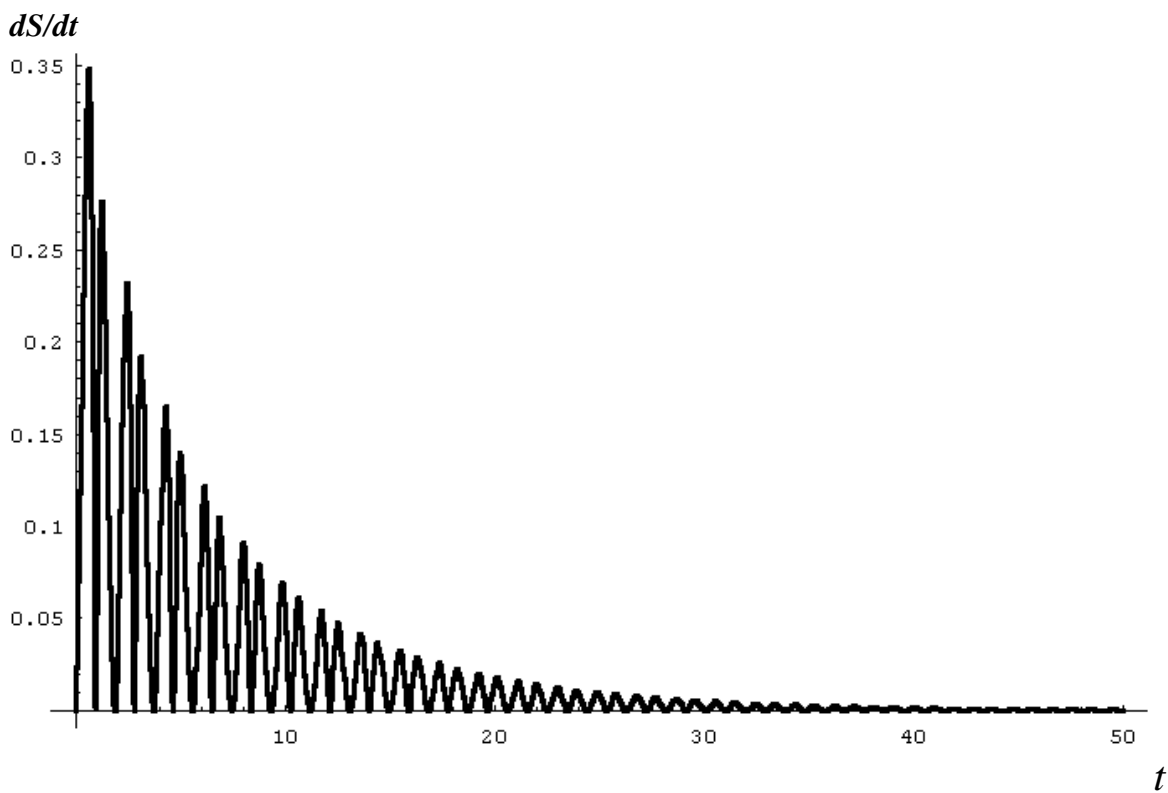

Fig. 4 *Representação da variação de entropia em ordem ao tempo em função do tempo. Note-se que à medida que o êmbolo fica mais lento a produção de entropia vai diminuindo.*

## Apêndice II

Considere-se 0,1 mole de um gás ideal clássico monoatómico para o qual $c_v = 1,5\ R$ (calor específico molar). O gás ocupa inicialmente o estado $A$ com o volume $V_A = 2,5\ 10^{-3}$ m$^3$ à temperatura $T_A = 300$ K e à pressão $p_A$ de $10^5$ Pa. A pressão devida ao peso de um êmbolo (que no início está bloqueado) é $p_e = 0,7\ 10^5$ Pa. Desbloqueado o êmbolo, este vai entrar em movimento dado a pressão $p_A$ ser superior a $p_e$. Admitindo que a transformação é adiabática só há troca de energia entre o gás e o êmbolo que muda de energia cinética e de energia potencial devido às colisões das partículas do gás. O estado final de equilíbrio, após algumas oscilações do êmbolo, pode ser facilmente calculado usando o princípio de conservação de energia e a igualdade das pressões do gás e do êmbolo, igualdade que se verifica no estado final de equilíbrio. Designemos por *B* este estado de equilíbrio final. Temos

$$\Delta U_{AB} + \Delta E_{AB} = 0 \qquad (1)$$

$$\Delta E_{AB} = \Delta Epot_{AB} = p_e(V_B - V_A) \qquad (2)$$

em que *U* é a energia do gás, *E* é a energia total do êmbolo, soma da energia cinética com a energia potencial - *Epot* é a energia potencial do êmbolo devida ao campo gravitacional que origina $p_e$. No estado *B*, dado o êmbolo estar em repouso, a energia cinética do êmbolo é também zero, como no estado *A*, e por isso a variação da energia cinética do êmbolo é zero. Dado

$$U = nc_V T$$

temos de (1) e (2)

$$-p_e(V_B - V_A) = nc_V(T_B - T_A) \qquad (3)$$

em que *n* = 0,1 mole.

A pressão do gás no ponto de equilíbrio final $p_B$ é igual à pressão do êmbolo $p_e$:

$$p_B = p_e = 0,7 \times 10^5 \, Pa = \frac{nRT_B}{V_B} = \frac{0,1 \times R \times T_B}{V_B}. \qquad (4)$$

(3) e (4) permitem determinar

$$V_B = 3,1 \times 10^{-3} m,^3 \qquad (5)$$
$$T_B = 264,5 \, K. \qquad (6)$$

O estado final *B* de repouso do êmbolo foi daquela forma calculado.
Voltemos agora a situar-nos no estado de repouso inicial *A* e libertemos o êmbolo. O êmbolo após ser desbloqueado vai atingir uma altura máxima que pode ser facilmente calculada se admitirmos que durante o troço da trajectória entre o ponto de equilíbrio *A* e o ponto correspondente à altura máxima que vamos designar por *C*, a pressão dinâmica sobre o êmbolo é aproximada pela pressão estática. Dado o êmbolo em *A* e *C* estar em repouso $\Delta Ecin = 0$ (*Ecin* é a energia cinética do êmbolo). Temos que

$$W = \int_{AC} -p'dV + \int_{AC} p_e dV = \Delta Ecin_{AC} = 0 \qquad (7)$$

e portanto

$$\int_{AC} - p'dV = \int_{AC} - p_e dV = -p_e(V_C - V_A). \qquad (8)$$

Como

$$\int_{AC} - p'dV = U_C - U_A \qquad (-p'dV = dU)$$

temos

$$U_C - U_A = -p_e(V_C - V_A) \qquad (9)$$

Dado

$$p' \cong p \qquad (dS \cong 0)$$

temos ( ver obtenção da eq. (28) em 3.)

$$p_C V_C^{\alpha+1} \cong p_A V_A^{\alpha+1}. \qquad (10)$$

A transformação é aproximadamente isentrópica. Podemos escrever (10) na forma

$$T_C V_C^{\alpha} \cong T_A V_A^{\alpha} \qquad (11)$$

em que $\alpha = 2/3$ se o gás for mono-atómico.

De (11)

$$T_C \cong \frac{T_A V_A^{\alpha}}{V_C^{\alpha}}. \qquad (12)$$

Atendendo a que

$$U_C - U_A \cong nR \times 1{,}5 \times (T_C - T_A) \qquad (13)$$

substituindo $U_C - U_A$ em (9) e tendo em conta (12) e os valores dados de $V_A$, $T_A$, $p_e$ e $n$, obtém-se:

$$V_C \cong 3{,}6 \times 10^{-3}\, m^3. \qquad (14)$$

Entre *A* e *C* verifica-se que

$$U_C - U_A = \int_{AC} -p'dV \cong \int_{AC} -pdV \qquad (15)$$

dado a pressão dinâmica ser aproximadamente a pressão estática, mas entre *A* e *B*

$$U_B - U_A = \int_{AB} -p'dV \neq \int_{AB} -pdV \qquad (16)$$

dado entre o estado inicial *A* e o estado final *B* não ser já aceitável ignorar o termo de variação da entropia, pelo que

$$U_B - U_A = \int_{AB} -p'dV = \int_{AB} (-pdV + TdS). \qquad (17)$$

De facto, entre *A* e *C* considerou-se, em boa aproximação, que a transformação é *quase-isentrópica*, o que já deixará de ser válido para transformações entre estados mais distanciados como *A* e *B*. Estas considerações são consistentes com os resultados da simulação no Apêndice I - no espaço de fase, no primeiro ciclo da trajectória de fase o sistema regressa a um estado muito próximo do estado inicial, descrevendo assim uma trajectória quase fechada correspondente a uma transformação quase-isentrópica; mas nos ciclos seguintes a trajectória vai-se progressivamente afastando do ciclo inicial, evidenciando assim uma variação significativa, e cumulativa, da entropia.

Desta forma compreende-se a razão da equação,

$$dU = -pdV + dQ \qquad (18)$$

poder conduzir a uma boa aproximação fazendo *dQ=0* (a transformação é dita adiabática) – é o que se verifica entre *A* e *C*, num pequeno troço da trajectória. Mas entre os estados inicial *A* e final *B* não se pode admitir a validade da equação com *dQ=0*, embora a transformação seja adiabática. De facto entre *A* e *B* temos que

$$dU = -p'dV = -pdV + TdS = dW + dQ \qquad (19)$$

com (as grandezas são aqui definidas formalmente)

$$dW = -pdV$$
$$dQ = TdS \neq 0$$

e, evidentemente, poderíamos escrever (mas agora também com significado físico)

$$dU = -p'dV = dW' + dQ' \qquad (20)$$

$$dQ' = 0 ............. (transformação..adiabática)$$
$$dW' = -p'dV ..... (transformação..não-isentrópica)$$

com

Em troços da trajectória (como entre *A* e *C*) poderemos ter aproximadamente

$$\int_{\acute{A}C} dW' \cong \int_{AC} dW \qquad \int_{\acute{A}C} dQ' \cong \int_{AC} dQ \qquad (21)$$

Mas entre *A* e *B*

$$\int_{\acute{A}B} dW' \neq \int_{AB} dW \qquad \int_{\acute{A}B} dQ' \neq \int_{AB} dQ \qquad (22)$$

Deste modo compreende-se como por uma via conceptualmente errada se podem obter resultados aproximadamente correctos. Evidentemente que através de uma via conceptualmente correcta podem-se obter os mesmos resultados, e obviamente outros quando as aproximações deixam de ser válidas – é o que se passa entre os estados inicial e final *A* e *B* em que deixa de ser válida a equação

$$dU = -pdV + dQ = dW + dQ$$

com *dQ=0*, devendo esta equação ser substituída pela equação (21). A equação é generalizada para transformações não "quase-estáticas" [6, 7].

Note-se que é fácil, através do modelo analisado, compreender a origem de diversas formulações da Termodinâmica: temos três "trabalhos" elementares – o da força peso, o da força estática e o da força dinâmica. Em transformações reversíveis estes trabalhos são iguais. Em transformações irreversíveis podem coincidir, aproximadamente, entre *A* e *C*, o trabalho da pressão dinâmica e da pressão estática - e o trabalho da pressão exterior, devido ao peso do êmbolo, é também igual a estes dois trabalhos entre esses dois pontos de equilíbrio do êmbolo. Obviamente que na equação do 1º Princípio não se pode arbitrariamente identificar um destes trabalhos com o termo *dW*, dado que, como vimos, é o trabalho da pressão dinâmica que, com generalidade (por exemplo ente *A* e *B*), permite verificar a Lei da Conservação da energia. A generalização da análise para uma parede adiabática móvel separando dois volumes de gás tem sido feita, mas tem originado controvérsia, não existindo ainda consenso sobre esta matéria. Esta última análise é importante ser referida, pois faz surgir relativamente ao problema anterior uma nova variável, a pressão do gás que agora passa a existir do outro lado do êmbolo. A pressão sobre o êmbolo devida às colisões das partículas, é agora a diferença das pressões exercidas em cada um dos lados do êmbolo e o trabalho da pressão dinâmica passa a ser o trabalho desta diferença de pressões. A razão da controvérsia resulta da atribuição de significado físico a uma grandeza *dQ* que, nesse contexto, não o tem, e que é considerada nula por a parede (o êmbolo) ser "adiabático" [7,12-26] (ver Abreu, R. http://xxx.lanl.gov/abs/cond-mat/0205566 ).